\documentclass[onecollarge,natbib]{svjour2}
\bibpunct{[}{]}{;}{n}{}{,} 
\smartqed  
\usepackage{graphicx,color}
\usepackage[normalem]{ulem}

\journalname{Few-Body Systems}
\begin{document}

\title{Scattering solutions  of Bethe-Salpeter equation in Minkowski and Euclidean spaces}


\author{J.~Carbonell     \and V.A.~Karmanov     
}


\institute{J.~Carbonell \at
              Institut de Physique Nucl\'eaire,
Universit\'e Paris-Sud, IN2P3-CNRS, 91406 Orsay Cedex, France\\
 \email{carbonell@ipno.in2p3.fr}  
\and  V.A.~Karmanov \at
              Lebedev Physical Institute, Leninsky Prospekt 53, 119991
Moscow, Russia \\
              \email{karmanov@sci.lebedev.ru}           
}

\date{Received: date / Accepted: date}

\maketitle

\begin{abstract}
We  shortly review  different methods to obtain the scattering solutions of the Bethe-Salpeter equation in Minkowski space.
We emphasize the possibility to obtain the zero energy observables in terms of the  Euclidean scattering amplitude.
\end{abstract}

\section{Introduction} \label{intro}

The Bethe-Salpeter (BS) equation \cite{bs} deals with  a -- preexisting -- Quantum Field Theory (QFT) object
defined as a vacuum expectation value of the T-ordered product of  Heisenberg field operators  \cite{GML_PR84_51}. 
For  the  two-body case in a scalar theory ($\phi$),   this object  has the expression
\begin{equation}\label{Phi_QFT}
\Phi(x_1,x_2;p) =  \langle  0 \mid T\left\{ {\phi}(x_1) \phi(x_2) \right\} \mid p \rangle ,
\end{equation}
where $\mid p\rangle$ is the state  vector  with total momentum $p$.

After removing some trivial dependence related to translational invariance, 
\[ \Phi(x_1,x_2; p)=\frac{1}{(2\pi)^{3/2}} \; {\Phi}(x; p) \;e^{-ip \cdot(x_1+x_2)/2}, \]
its Fourier transform 
\[  \Phi(x ; p)= \int  d^4x  \; \Phi(k ; p)\; e^{-ik\cdot x}  \qquad x=x_1-x_2 \]
defines the momentum space BS amplitude.
 Bethe and  Salpeter wrote a covariant 4D integral equation for the amplitude $\Phi(k ; p)$, which for the two-body bound state reads
 \begin{equation}\label{BS_BS}
\Phi(k;p) = S_1(k,p)\; S_2(k,p) \;  \int{d^4k'\over(2\pi)^4} \; iK(k,k';P)\; \Phi(k' ; p)  
\end{equation}
where
\begin{equation}
S_1(k,p) = \frac{i} {\left({p\over2}+k\right)^2-m^2+i\epsilon}  \qquad
S_2(k,p) = \frac{i} {\left({p\over2}-k\right)^2-m^2+i\epsilon} 
\end{equation}
are the free constituent propagators and $iK$ the interaction kernel.
If  the propagators were dressed and the kernel  was a sum of all the irreducible Feynman graphs of the theory, solving (\ref{BS_BS}) would be
equivalent to solve the full QFT.
This is however a wishful thinking. In practice only very poor representations
of the interaction are used: {most of nuclear physics calculations use ladder approximation and free propagators, though
 dressed propagators are systematically used in the Schwinger-Dyson approach \cite{GE_LCM_2015}.}

There could be some misunderstanding on what  BS amplitude denotes: either the quantity (\ref{Phi_QFT}) computed whit  the full
QFT solution  or the quantity $\Phi$ obtained by  solving (\ref{BS_BS})  with a -- necessarily -- truncated interaction kernel.
These quantities can strongly differ from each other, but we  live with it.

Besides the bound states equation (\ref{BS_BS}), we have also considered the corresponding scattering equation for the  process $k_{1s}+k_{2s}\to k_1+k_2$. 
The two-body scattering amplitude $F(k;p)$  satisfies an inhomogeneous BS equation which, for  spinless particles in Minkowski space, reads:
\begin{equation}\label{BSE}
F(k;p)=K(k,k_s)- i\int\frac{d^4k'}{(2\pi)^4} \frac{K(k,k') F(k';p)}
{\left[\left(\frac{p}{2}+k'\right)^2-m^2+i\epsilon\right]  \left[\left(\frac{p}{2}-k'\right)^2-m^2+i\epsilon\right]} 
\end{equation}
with $2k=k_1-k_2$ and $p=k_1+k_2$.

This equation is plagued with singularities:
{\it (i)} in the two-body propagators,
{\it (ii)} in most of the interaction kernels even the simplest ones,
{\it (iii)} in the (inhomogeneous) Born term
and
{\it (iv)} in the amplitude $F$ itself, being difficult to represent it in terms of smooth functions.

To avoid these troubles -- and get some bound state solutions  -- Wick proposed \cite{WICK_54}
to  introduce a new variable $k_0=ik_4$ in terms of wich the 4D  Minkowski metric  is changed into an Euclidean one:
\[ k^2 = k_0^2 - \vec{k}^2  = -(k_4^2+\vec{k}^2) \equiv -k_E^2 \]
This change of metric leads to a smooth integral equation  for the Euclidean amplitude, related to the Minkowski one by
\begin{equation}\label{F_E}
 F_E (k_4,\vec{k})= F_M(ik_4,\vec{k}) 
\end{equation}
which is easily  solvable by standard methods.
Until very recently most of the existing solutions were found in this way.
The are however some problems in performing the so called Wick rotation:
\begin{enumerate}
\item It is not a simple change of variable but underlies an application of the Cauchy theorem with two different integration contours. 
The equivalence of  both theories -- Minkowski and Euclidean -- requires a careful analysis of the singularities crossed when moving from one to other. 
For the bound state case, the singularities come only in  the interaction kernel and the validity has been proved only for simple cases.
For the scattering case, they come also in the propagators and the corresponding residues must be taken into account.
\item The total mass $M$ of the system is invariant under such transformation but the amplitude not. In practice it is very difficult -- if possible at all --
to recover $\Phi_M$ from $\Phi_E$ (see   contribution  \cite{Tobias_LCM_2015})
\item Other observable --  scattering amplitudes, form factors -- are defined and computable in terms of Minkowski amplitudes.
It is worth noticing, however, that the distinction between Euclidean and Minkowski amplitues is meaningful only when 
keeping for both of them real arguments $(k_0,\vec{k})$  and $(k_4,\vec{k})$ and the later statement must be understood in this conventional sense.  
Euclidean amplitudes with imaginary arguments  \cite{GE_LCM_2015} are  equivalent to Minkowski ones and can  lead to correct results.
\end{enumerate}
All these reasons motivated several authors  to solve the BS equation in Minkowski  space,
first for the bound state   \cite{Kusaka,bs1-2,Sauli_JPG08,Carbonell:2010zw,fsv-2012} and later for the scattering problem
 \cite{CK_PLB_2013,bs_long,fsv-2014,fsv-2015}.
Obtaining these numerical solution  is  indeed quite difficult  and was not tried until very recently.

\section{Minkowski space solutions: Light-Front projection and Nakanishi representation}\label{sec1}

Some years ago we proposed \cite{bs1-2} a method based on an integral transform of the original BS equation which make all things regular.
It was inspired by the relation between the Light-Front (LF) wave function $\Psi$  and the BS amplitude $\Phi$
\[Ã\Psi(k_1,k_2,p,\omega) =  \frac{ (\omega\cdot k_1)(\omega\cdot k_2)}{\pi(\omega\cdot p)} \int_{-\infty}^{+\infty}  d\beta \; \Phi(k+\beta\omega) \]
This integral transform, denoted symbollicaly by $\Psi=L\Phi$,  maps the singular BS amplitude into a smooth wave function $\Psi$ and was applied to the  full BS equation (\ref{BS_BS})
according to
\begin{equation}\label{Formal}
\Phi =  G_0 K \Phi \qquad  \Longrightarrow\qquad   L \Phi = L G_0 K \Phi  \qquad  \Longrightarrow\qquad  \Psi= L G_0 K \Phi
\end{equation} 
In order to use the same unknown function on both sides of (\ref{Formal})   and avoid the inversion of $L$, we have used the Nakanishi representation \cite{nak63}  of the BS amplitude
\begin{equation}\label{Naka}
\Phi(k,p)=  \int_{-1}^{+1} dz \int_0^{\infty} d\gamma  \;  {1\over [  \gamma+ m^2 -  {M^2\over 4} - k^2 - p\cdot k z - i \epsilon]^3 }    \; g(\gamma,z)  
\end{equation} 
and obtained an integral equation for the Nakanishi weight function $g$, on the form:
\begin{equation}\label{LFN_BS}
ÃLNg = (LG_0KN)g 
\end{equation}
The details of this derivation and the analytic expressions of the kernels can be found in \cite{bs1-2}.

The function $g$ is metric-independent and finite although it can display -- at least for zero energy scattering  -- Heaviside-like discontinuities \cite{fsv-2015,VS_LCM_2015}.
All the Minkowski space poles of $\Phi$ are contained in the denominator  of  (\ref{Naka}).
Once $g$ is known, the BS amplitude -- either in Minkowski or in Euclidean space --  can be computed by means of (\ref{Naka}). 
A similar, one-dimensional, expression exists for the Light-Front wave function $\Psi$ in terms of $g$ \cite{bs1-2}
\begin{equation}\label{Psi_g} 
\Psi(\gamma,z)=\frac{1-z^2}{4}\int_0^{\infty} \;d\gamma'\; \frac{g(\gamma',   z )}{ \left[  \gamma  +m^2  - {M^2\over4}  (1-z^2)  + \gamma '\right]^2}  
 \end{equation}
These two possibilites make the Nakanishi weight function $g$ especialy well adapted for establishing a relation  between Euclidean and Minkowski BS solutions. 
Indeed, if $g$ could be determined  either from the LF wave function  -- inverting (\ref{Psi_g}) -- or  from an Euclidean BS solution -- inverting (\ref{Naka}) -- it could be used
to computed the Minkowski amplitude and related observables.
Care must be taken however  \cite{Tobias_LCM_2015} with the ill-defined character in  the numerical inversion of  (\ref{Naka}) and (\ref{Psi_g}).

Further applications  of this LF/Nakanishi approach have been developed in the recent years.
\begin{itemize}
\item The solutions of  (\ref{LFN_BS}) were used to  compute the electromagnetic form factors \cite{Carbonell:2008tz,ck-trento} in terms of $g$.   
\item Equation (\ref{LFN_BS}) was generalized to the two-fermion system. The usual  one-boson exchange kernels 
have been dervied and the bound state solution for the $J^{\pi}=0^+$ states with the Yukawa coupling  have been obtained in \cite{Carbonell:2010zw,Wayne_LCM_2015}.
\item In  \cite{fsv-2012},  equation (\ref{LFN_BS}) was written in the form $ g=  K'\; g$.
This   non trivial analytic inversion of the left-hand-side kernel of  (\ref{LFN_BS}),  allows to compute unambiguously the function $g$ and study its analytic properties.
The bound state solutions found by solving this equation were in perfect agreement with the preceding results.
The equation was extended to the scattering states and the scattering length could be accurately computed \cite{fsv-2015}.
\end{itemize}

\section{Minkowski space solutions: direct approach}\label{sec2}

A second method was developed in \cite{CK_PLB_2013,bs_long} aiming to directly solve the original equation without any transform and  taking  into account all the singularities  of  equation (\ref{BSE}). 
This direct approach has proved its efficiency  both for the bound and, especially, for  the scattering states. 
It provided for the first time the full off-shell scattering amplitude at the price of
a careful analytical work to transform all the singular terms, from the kernels as well as from the propagators, into a smooth integrands.

For instance, the BS equation for the S-wave can be written in the following form
\begin{small}
\begin{eqnarray}
F_0(k_0,k)  &=& F^B_{0}(k_0,k) \cr
  &  +&  \frac{i\pi^2 k_s}{8\varepsilon_{k_s}} W_0^S(k_0,k,0,k_s) F_0(0,k_s)
 \cr
                 &+& \frac{\pi}{2M} \int_0^{\infty}  \frac{dk'}{ \varepsilon_{k'} ( 2\varepsilon_{k'}-M) }    \left[{k'}^2 W_0^S(k_0,k, a_-,k') F_0(| a_- |,k')
     -\frac{2 {k_s}^2\varepsilon_{k'}}{\varepsilon_{k'} +\varepsilon_{k_s}}W_0^S(k_0,k,0,k_s) F_0(0,k_s)\right]
\cr
                    &-&    \frac{\pi}{2M} \int_0^{\infty} \frac{{k'}^2 dk'}{\varepsilon_{k'} (2\varepsilon_{k'}+M) }      W_0^S(k_0,k, a_+,k') F_0(a_+,k' )   \cr
                 &+&  \frac{i}{2M}  \int_0^{\infty}  \frac{{k'}^2dk' }{\varepsilon_{k'}}   \int_0^{\infty} dk'_0 \left[ \frac{ W^S_{0}(k_0,k,k'_0,k') F_0(k'_0,k')  - W^S_{0}(k_0,k,a_-,k')  F_0(|a_- |,k')}{  {k'}_0^2-a_-^2 }\right]  \cr
                &-&    \frac{i}{2M}  \int_0^{\infty} \frac{{k'}^2dk'} {\varepsilon_{k'}}     \int_0^{\infty} dk'_0 \left[ \frac{ W^S_{0}(k_0,k,k'_0,k') F_0(k'_0,k')  - W^S_{0}(k_0,k,a_+,k') F_0(a_+,k')}  {{k'}_0^2-a_+^2 } \right]  \label{EQD}
\end{eqnarray}
\end{small}
where $W^S_0$ is the S-wave kernels and $F^B_{0}$ the corresponding Born term. 
Notice, apart from the inhomogeneous term, the existence of zero-, one- and two-dimensional integral terms.
The details of the derivation and some illustrative results  can be found in \cite{bs_long}.

Having computed the scattering solutions, the corresponding "bound $\to$ scattering state" transition electromagnetic form factor was found in \cite{tff_long}.

\section{Deriving an Euclidean equation}\label{sec3}

In contrast to the bound state case, the scattering BS equation  cannot be in general transformed into an equation for  the Euclidean amplitude $F_E$ alone.
Indeed, let us consider  equation (\ref{BSE})
and assume we want to obtain from it an equation for the Euclidean amplitude $F_E$ defined in (\ref{F_E}).
In order to  properly apply the Wick rotation, we have to take into  account  the pole singularities associated with the propagators in (\ref{BSE}) which are given by 
\begin{eqnarray}
{k'}^{(1)}_0(k,k_s)&=&\phantom{-}\varepsilon_{k_s}+\varepsilon_{k'} - i\epsilon = +a_+  -   i\epsilon            \cr
{k'}^{(2)}_0(k,k_s)&=&\phantom{-}\varepsilon_{k_s}-\varepsilon_{k'}+ i\epsilon  = -a_-    +  i\epsilon      \cr
{k'}^{(3)}_0(k,k_s)&=&-\varepsilon_{k_s}+\varepsilon_{k'}- i\epsilon  = +a_- - i\epsilon      \cr
{k'}^{(4)}_0(k,k_s)&=&-\varepsilon_{k_s}-\varepsilon_{k'}+ i\epsilon  = - a_+ + i\epsilon   \label{Poles}
\end{eqnarray}
with
$a_{\pm}(k',k_s) =\varepsilon_{k'}   \pm \varepsilon_{k_s} $

\begin{figure}[htbp]
\centering\includegraphics[width=7.5cm]{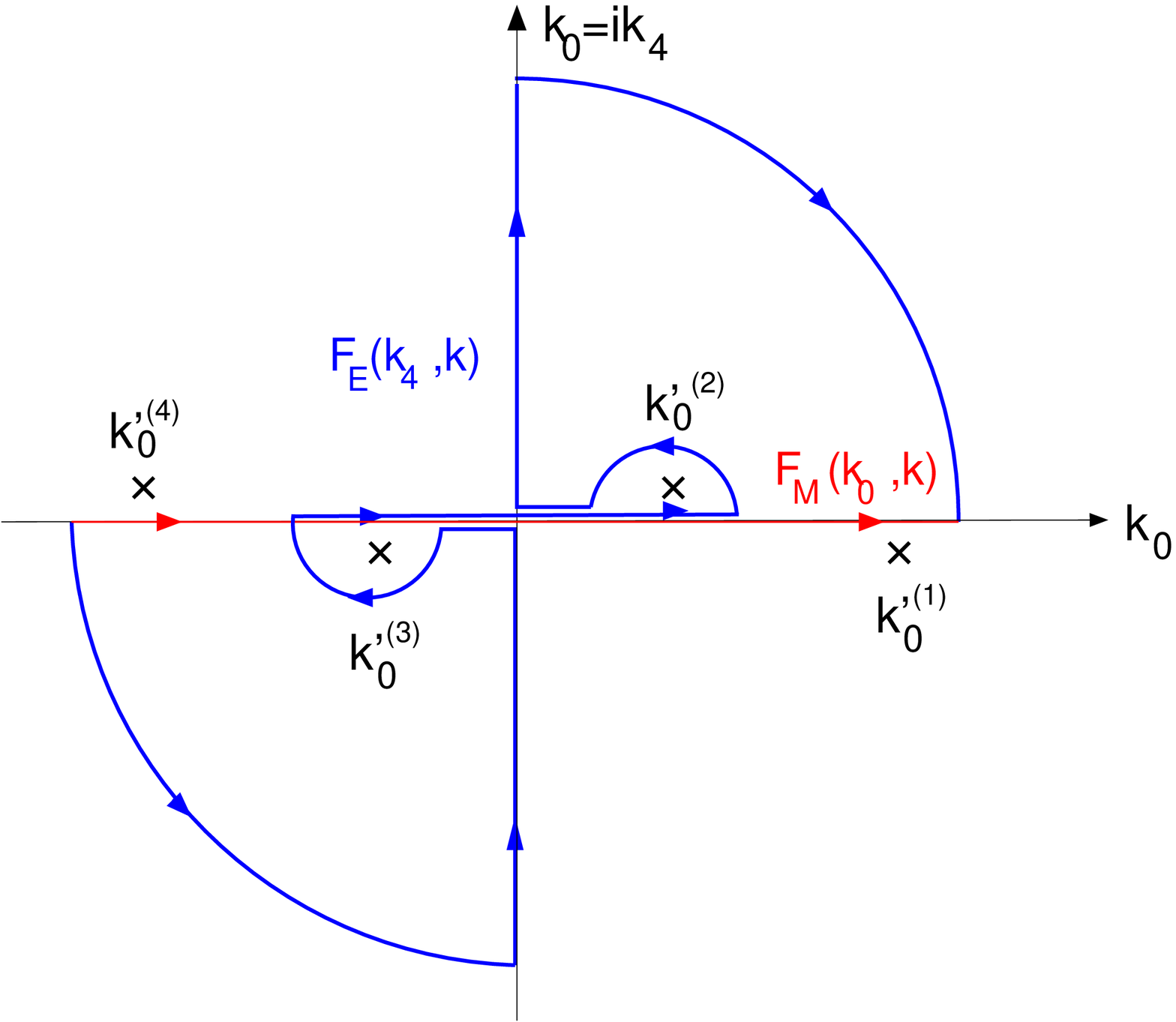}
\caption{Singularities of propagators for scattering state and  integration contour in  complex $k_0$ plane ($k'<k_s$).}\label{fig1a}
\end{figure}

In  the case $k'<k_s$, their positions in the complex  $k_0$-plane are shown in Fig. \ref{fig1a}. 
When the integration contour is rotated,  the   singularities 
${k'}^{(2)}_0$ and ${k'}^{(3)}_0$ are crossed and the corresponding residues of the integrand at these poles will contribute to the integral term.

Equation (\ref{BSE})  is then transformed into:
\begin{equation}\label{BSEc}
F^E(k_4,\vec{k};\vec{k}_s)=V^B(k_4,\vec{k};\vec{k}_s)+ \int\frac{d^4k'}{(2\pi)^4} \frac{V(k_4,\vec{k'};k'_4,\vec{k'}) F^E(k'_4,\vec{k'};\vec{k}_s)} {({k'_4}^2 +a_-^2)({k'_4}^2+a_+^2)} + S(k_4,k,k_s)
\end{equation}
where $V$  and $V^B$ are respectively the Wick-rotated interaction  kernel and Born term.

We emphasize that the obtained equation does not simply follow from the original one after  the variable  change  $k_0\to ik_4$. 
The additional term $S$ is the contribution of the two  poles ${k'}^{(2)}_0$ and ${k'}^{(3)}_0$ displayed in Fig. \ref{fig1a}. 
It can be shown \cite{bs_long,CK_PLB_2016} that $S$ depends on the Minkowski amplitude at the particular value of its  $k_0$ argument
\[F_M(k_0= \varepsilon_{k_s} -\varepsilon_{k},k)   \]
Consequently, when rotating the initial BS scattering equation, the Euclidean and Minkowski amplitudes are  in general coupled and one cannot obtain a scattering equation
involving $F_E$ alone.
 
A second integral equation is  required  and
after some algebraic manipulations \cite{bs_long} one ends with a system of two coupled integral equations involving both $F_M$ and $F_M$ and denoted simbolically
\begin{eqnarray}
F_E(k_4,\vec{k} )                                                           &=& {\cal I}_1 [  F_E(k'_4,\vec{k}'), F_M(\varepsilon_{k_s} - \varepsilon_{k'},\vec{k}') ]  \cr
F_M(\varepsilon_{k_s} - \varepsilon_{k} ,\vec{k})  &=& {\cal I}_2 [  F_E(k'_4,\vec{k}'), F_M(\varepsilon_{k_s} - \varepsilon_{k'},\vec{k}') ]  \label{2CEQ}
\end{eqnarray}
This system of equations -- whose Minkowski part remains singular  -- was first derived in \cite{tjon}, re-derived and solved in \cite{bs_long} to check our direct Minkowski solution $F_M$
at the particular value  $k_0=\varepsilon_{k_s} - \varepsilon_{k}$.
The solution of (\ref{2CEQ}) provides $F_E$ in the full domain  but
$F_M$ is limited to a restricted manifold  and does not provide the full solution of the scattering problem.
However, on mass-shell $k_0=k_4=0$  and since  $F_M(0,k)=F_E(0,k)$,  the Euclidean amplitude computed this way can be also used to obtain the phase shifts.
The results provided by our direct method (\ref{EQD}) are in agreement with the solution of (\ref{2CEQ}) and with those of  \cite{tjon}.

\subsection{The case of zero energy scattering}\label{sec4}

Of particular interest is the zero energy scattering.
It can be shown  \cite{CK_PLB_2016} that in the limit $k_s\to0$ 
the term $S$ in Eq. (\ref{BSEc}) vanishes and the four-dimensional BS equation for the Euclidean amplitude in the zero energy limit takes the form
\begin{equation}\label{BSEc0}
F^E(k_4,\vec{k};0)=V^B(k_4,\vec{k};0)+ \int\frac{d^4k'}{(2\pi)^4} \frac{V(k_4,\vec{k'};k'_4,\vec{k'}) } {({k'_4}^2 +a_-^2)({k'_4}^2+a_+^2)} \; F^E(k'_4,\vec{k'};0)
\end{equation}
This is a self-consistent integral equation for determining $F^E(k_4,\vec{k};\vec{k}_s=0)$ without any coupling to the Minkowski amplitude $F$.
This  decoupling is valid only in the limit $k_s\to 0$.
 
In the  S-wave case,  Eq. (\ref{BSEc0}) reduces to the following two-dimensional integral equation:
\begin{equation}\label{eq1}
F_0(k_4,k)=V_0^B(k_4,k) +\int_0^{\infty}{k'}^2dk'\int_0^{\infty}dk'_4\frac{V_0(k_4,k;k'_4,k')}{({k'_4}^2+{a'_-}^2)({k'_4}^2+ {a'_+}^2)}\;F_0(k'_4,k')
\end{equation}
with $ a'_\pm=\sqrt{m^2+{k'}^2}\pm m$ and where we have omitted the superscript on $F^E$.
This equation determines the S-wave zero energy half-mass shell scattering amplitude $F_0$.
The scattering length is given by
\begin{equation}\label{eq3}
a_0=-\frac{1}{m}F_0(k_4=0,k=0)
\end{equation}
A more detailed derivation  can be found in \cite{CK_PLB_2016}.

The interest in deriving equation (\ref{eq1})    is not only a significant simplification in  computing the BS scattering length 
but in the fact that this fundamental quantity can be obtained from a purely Euclidean solution. 
Although considered here in the approximate framework of BS equation, the "exact"  Euclidean amplitude (\ref{Phi_QFT}) --
solution of the  full QFT problem --  is nowadays accessible  in the Lattice approach.
It is the basic ingredient of the HAL-QCD  collaboration  to obtain the  {\it ab initio} Nucleon-Nucleon potential  from QCD
 \cite{Ishii:2006ec} but  it has never been used to extract the scattering observables. 

The possibility to compute a scattering amplitude  from Euclidean correlators in infinite space 
is forbidden by the  Maiani-Testa  theorem \cite{Maiani_Testa_PLB245_1990}. This result
is in agreement with the impossibility discussed above to obtain BS scattering observable without a coupling
to the Minkowski amplitudes, but does not apply to zero scattering energy where the phase shifts vanish.

Luscher et al. \cite{ML_CMP104_86} circumvented this problem and
proposed a method to compute the phaseshifts based on the $V$-dependence  of the 2-particle energies confined in a box with periodic boundary conditions.
This approach was  successfully applied  to several hadronic systems in {\it ab initio}  QCD  \cite{Detmold:2015jda}. 

The use of  Eq. (\ref{eq3}) constitutes  an alternative method to compute the scattering length  in the Lattice calculations,
directly from the Fourier transform  of the Euclidean  BS amplitude defined in  (\ref{Phi_QFT}).
In the  framework of BS equation, such a possibility is justified by the existence of Eq. (\ref{eq1}) allowing to compute the Euclidean BS amplitude in a purely Euclidean formalism  as in the Lattice approach.

The numerical solutions of Eq.  (\ref{eq1}) with one-boson exchange  kernel were obtained in \cite{CK_PLB_2016}.
The scattering  length $a_0$ was extracted  from  the Euclidean amplitude (\ref{eq3}) at the origin.
The value is in  full agreement with the results of our previous work \cite{CK_PLB_2013}  and from those of Ref. \cite{fsv-2015}, obtained by independent methods. 

\begin{figure}[htbp]
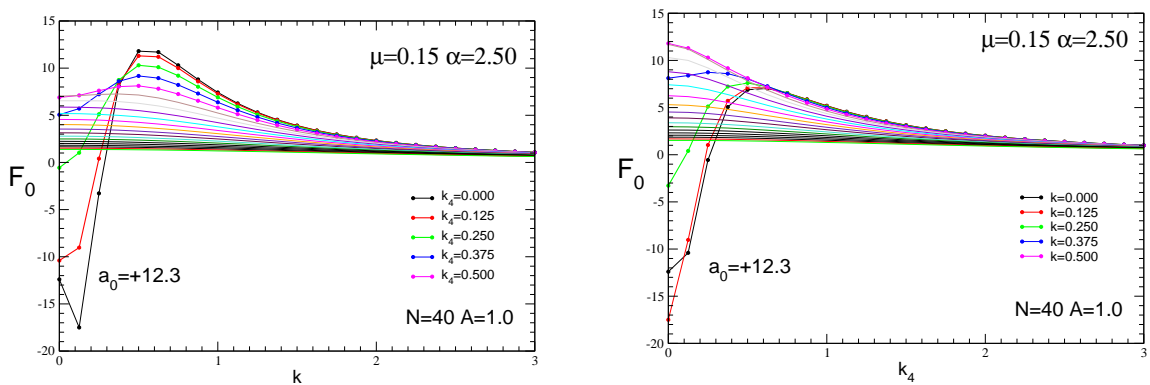

\centering\includegraphics[width=7.cm]{Fk.eps}\hspace{0.9cm}
\centering\includegraphics[width=7.cm]{Fk4.eps}
\caption{Euclidean scattering amplitude $F_0(k_4,k)$ as a function of $k$ for different values of  $k_4$ (left) and as a function of  $k_4$  for different values of $k$ (right). It corresponds to $m=1$, $\mu=0.15$ and $\alpha=2.5$.
The  scattering length value is given by $a_0=-F(0,0)=12.3$.}\label{F_mu_0.15_a_2.5}   
\end{figure}

Figure \ref{F_mu_0.15_a_2.5} displays the amplitude $F_0(k_4,k)$ as a function of $k$ at fixed values of $k_4$ (left panel) 
and as a function of $k_4$  at fixed values of $k$ (right panel) for the parameters $m=1$, $\mu=0.15$ and $\alpha=2.5$.
The scattering length $a_0=12.3$ is directly readable in both panels.
For these parameters  the two-body system has two bound states, which are responsible for the structure in the  amplitude.

\vspace{.5cm}

\vspace{-0.cm}
\section{Conclusion}\label{concl}

We  reviewed  different methods to obtain the scattering solutions of the Bethe-Salpeter equation in Minkowski space.

We showed that the Bethe-Salpeter  scattering amplitude  at  zero  energy obeys a purely Euclidean equation, as it is the case for the bound states.
This decoupling between Euclidean and Minkowski BS amplitudes is  only possible for zero energy scattering   and allows determining the  scattering length from the Euclidean BS amplitude.

Such a possibility  suggests to extract the scattering length in Lattice  calculations from a direct computation of the Euclidean
Bethe-Salpeter amplitude  (\ref{Phi_QFT}) in momentum space  and provides an alternative  to the Luscher method. 


\end{document}